\documentclass[]{article}

\usepackage{amsmath,amssymb,amsthm} % AMS packages
\usepackage{mathrsfs,mathtools} % additional math packages

\usepackage{hyperref}
\hypersetup{colorlinks,urlcolor=blue}
\hypersetup{
    colorlinks,
    linkcolor={blue},
    citecolor={blue},
    urlcolor={blue}
}

\newtheorem{definition}{Definition}

\DeclareMathOperator*{\argmax}{arg\,max}

\title{Photon Counting Accuracy in Digital Image Sensors}
\author{Aaron J.~Hendrickson\\ ajh4184@gmail.com}

\begin{document}

\maketitle

\begin{abstract}
A statistical model for data emanating from digital image sensors is developed and used to define a notion of the system level photon counting accuracy given a specified quantization strategy.  The photon counting accuracy for three example quantization rules is derived and the performance of each rule is compared.
\end{abstract}

\begin{table}[b]\footnotesize\hrule\vspace{1mm}
	Keywords: photon counting, sub-electron read noise, quanta image sensor, quantization.\\
\end{table}

%%%%%%%%%%%%%%%%%%%%%%%%%%%%%%%%%%%%%%%%%%%%%%%%%%%%%%%%%%%%%%%%%%%%%%%%%%%%%%%%%%%%
%%%%%%%%%%%%%%%%%%%%%%%%%%%%%%%%%%%%%%%%%%%%%%%%%%%%%%%%%%%%%%%%%%%%%%%%%%%%%%%%%%%%
%%%%%%%%%%%%%%%%%%%%%%%%%%%%%%%%%%%%%%%%%%%%%%%%%%%%%%%%%%%%%%%%%%%%%%%%%%%%%%%%%%%%

\section{Introduction}

In recent years, the advent of Deep Sub-Electron Read Noise (DSERN) technology has brought with it the ability to turn image sensors into accurate photon counting (photon number resolving) devices \cite{9768129}. In a typical image formation chain, photons are converted into charge carriers within a pixel, e.g.~electrons, which in turn are sensed as a voltage and finally quantized via an Analog-to-Digital Converter (ADC). By carefully placing the threshold voltages in the ADC, the quantized signal can be directly interpreted as an estimate of the number of charge carriers generated in the pixel at the beginning of the image formation chain. As such, the accuracy of the estimated number of charge carriers depends not only on the noise present in the sensor, but also the placement of the ADC voltage thresholds \cite{Fossum_BER_paper,threshold_calib,yin2021threshold}. Here, we provide a natural definition of Photon Counting Accuracy (PCA) and study how the choice of quantization strategy (placement and number of voltage thresholds) affects the overall counting accuracy.

%%%%%%%%%%%%%%%%%%%%%%%%%%%%%%%%%%%%%%%%%%%%%%%%%%%%%%%%%%%%%%%%%%%%%%%%%%%%%%%%%%%%
%%%%%%%%%%%%%%%%%%%%%%%%%%%%%%%%%%%%%%%%%%%%%%%%%%%%%%%%%%%%%%%%%%%%%%%%%%%%%%%%%%%%
%%%%%%%%%%%%%%%%%%%%%%%%%%%%%%%%%%%%%%%%%%%%%%%%%%%%%%%%%%%%%%%%%%%%%%%%%%%%%%%%%%%%

\section{Statistical Model}

Consider a pixel that generates an average of $H\, (e\text{-})$ charge carriers (photoelectrons) during the integration period and assume the number of photoelectrons generated, $K$, is distributed according to the Poisson distribution $K\sim\operatorname{Poisson}(H)$. During read-out, the accumulated charge, $Q=qK$, where $q$ is the charge of an electron, is placed on the capacitance, $C$, of a source follower transistor. Ideally the voltage at the output of the source follower would be $U=Q/C=qK/C$, but due to thermal (Johnson-Nyquist) read noise, is corrupted by Gaussian read noise. Thus, the measured voltage is described by
\begin{equation}
U=sK+R_v+\mu,
\end{equation}
where $s=q/C$ is the charge detection sensitivity in $(\mu\mathrm V/e\text{-})$, $K\sim\operatorname{Poisson}(H)$ with quanta exposure $H\,(e\text{-})$, $R_v\sim\mathcal N(0,\sigma_v^2)$ with read noise $\sigma_v\,(\mu\mathrm V)$, and DC offset $\mu$ in $(\mu\mathrm V)$.

For what follows we define the corrupted voltage signal as \cite{hendrickson_2023_PCHEM_theory,hendrickson_2023_PCHEM_verification}
\begin{equation}
U=\frac{K+R}{g}+\mu,
\end{equation}
where $g=1/s$ is the conversion gain in $(e\text{-}/\mu\mathrm V)$ and $R\sim\mathcal N(0,\sigma^2)$ with read noise $\sigma=\sigma_v/s$ in $(e\text{-})$.

Upon inspection, the density of the signal voltage given the electron number is normally distributed as $U|K=k\sim\mathcal N(\mu+k/g,(\sigma/g)^2)$, so that the density of $U$ is
\begin{equation}
    f_U(u)=\sum_{k=0}^\infty\mathsf P(K=k)f_{U|K}(u|k)=\sum_{k=0}^\infty\frac{e^{-H}H^k}{k!}\frac{g}{\sigma}\phi\left(\frac{u-(\mu+k/g)}{\sigma/g}\right),
\end{equation}
where $\phi(x)\coloneqq\frac{1}{\sqrt{2\pi}}e^{-x^2/2}$ is the standard normal Gaussian probability density.  In this work we will also make use of the standard normal distribution
\begin{equation}
    \Phi(x)\coloneqq\int_{-\infty}^x\phi(t)\,\mathrm dt=\frac{1}{2}(\operatorname{erf}(x/\sqrt 2)+1),
\end{equation}
where $\operatorname{erf}(x)$ is the error function.

%%%%%%%%%%%%%%%%%%%%%%%%%%%%%%%%%%%%%%%%%%%%%%%%%%%%%%%%%%%%%%%%%%%%%%%%%%%%%%%%%%%%
%%%%%%%%%%%%%%%%%%%%%%%%%%%%%%%%%%%%%%%%%%%%%%%%%%%%%%%%%%%%%%%%%%%%%%%%%%%%%%%%%%%%
%%%%%%%%%%%%%%%%%%%%%%%%%%%%%%%%%%%%%%%%%%%%%%%%%%%%%%%%%%%%%%%%%%%%%%%%%%%%%%%%%%%%

\section{PCA Definition}

With the signal voltage $U$, the goal is to now quantize $U$ back into an integer which represents an estimate, $\tilde K$, of the photoelectron number $K$. The resulting mapping (quantization) induces a partition of the real line $\Pi=(\Pi_0,\Pi_1,\dots)$ such that $\cup_k\Pi_k=\mathbb R$. This quantization can thus be defined as a mapping of $U$ to integers via
\begin{equation}
    \tilde K(U)=k\iff U\in\Pi_k.
\end{equation}

With this general description of quantization at hand, a natural definition of PCA follows.

\begin{definition}[Photon Counting Accuracy]
\label{def:PCA}
    Let $\tilde K(U)$ denote a quantization rule that induces a partition $\Pi$ on the real line. The photon counting accuracy for the induced partition is then defined as
    \begin{equation}
        \mathsf{PCA}(\Pi)\coloneqq\mathsf P(\tilde K=K).
    \end{equation}
\end{definition}

Definition \ref{def:PCA} provides a very general but natural description of PCA as the consequence of a chosen partition $\Pi$. To derive an explicit expression for the PCA, the law of total probability is utilized to write
\begin{equation}
    \begin{aligned}
        \mathsf{PCA}(\Pi)
        &=\mathsf P(\tilde K=K)\\
        &=\mathsf E(\mathsf P(\tilde K=K|K=k))\\
        &=\sum_{k=0}^\infty\frac{e^{-H}H^k}{k!}\mathsf P(U\in\Pi_k|K=k).
    \end{aligned}
\end{equation}
Given that $U|K=k\sim\mathcal N(\mu+k/g,(\sigma/g)^2)$ the final form is realized as
\begin{equation}
    \mathsf{PCA}(\Pi)
    =\sum_{k=0}^\infty\frac{e^{-H}H^k}{k!}\int_{\Pi_k}\frac{g}{\sigma}\phi\left(\frac{u-(\mu+k/g)}{\sigma/g}\right)\,\mathrm du.
\end{equation}

%%%%%%%%%%%%%%%%%%%%%%%%%%%%%%%%%%%%%%%%%%%%%%%%%%%%%%%%%%%%%%%%%%%%%%%%%%%%%%%%%%%%
%%%%%%%%%%%%%%%%%%%%%%%%%%%%%%%%%%%%%%%%%%%%%%%%%%%%%%%%%%%%%%%%%%%%%%%%%%%%%%%%%%%%
%%%%%%%%%%%%%%%%%%%%%%%%%%%%%%%%%%%%%%%%%%%%%%%%%%%%%%%%%%%%%%%%%%%%%%%%%%%%%%%%%%%%

\section{Examples}

Now equipped with Definition \ref{def:PCA}, this section introduces three examples of quantization strategies and the resulting PCA is evaluated.

%%%%%%%%%%%%%%%%%%%%%%%%%%%%%%%%%%%%%%%%%%%%%%%%%%%%%%%%%%%%%%%%%%%%%%%%%%%%%%%%%%%%

\subsection{One-Bit Quantizer}

Consider the one-bit quantizer
\begin{equation}
    \tilde K_\star(U)=
    \begin{cases}
        0, &U\in(-\infty,\mu+\frac{1}{2g}]\\
        1, &\text{otherwise},
    \end{cases}
\end{equation}
which induces the partition $\Pi_\star$. Working from Definition \ref{def:PCA}, the PCA is wirtten as
\begin{equation}
    \mathsf{PCA}(\Pi_\star)=e^{-H}\int_{-\infty}^{\mu+\frac{1}{2g}}\frac{g}{\sigma}\phi\left(\frac{u-\mu}{\sigma/g}\right)\,\mathrm du+e^{-H}H\int_{\mu+\frac{1}{2g}}^\infty\frac{g}{\sigma}\phi\left(\frac{u-(\mu+1/g)}{\sigma/g}\right)\,\mathrm du.
\end{equation}
Making use of the standard normal distribution property $\Phi(x)=1-\Phi(-x)$, one has after some algebraic manipulations
\begin{equation}
    \mathsf{PCA}(\Pi_\ast)=e^{-H}(H+1)\Phi\left(\frac{1}{2\sigma}\right).
\end{equation}
In the limit of zero read noise one finds $\lim_{\sigma\to 0^+}\mathsf{PCA}(\Pi_\star)=e^{-H}(H+1)$, which is less than one (except when $H=0$).  The fact that the PCA is less than unity even in the limit of zero read noise is a direct consequence of the finite bit-depth of the quantizer.

%%%%%%%%%%%%%%%%%%%%%%%%%%%%%%%%%%%%%%%%%%%%%%%%%%%%%%%%%%%%%%%%%%%%%%%%%%%%%%%%%%%%

\subsection{Infinite-Bit Uniform Quantizer}

Now consider the infinite-bit uniform quantizer
\begin{equation}
    \tilde K_\ast(U)=
    \begin{cases}
        0, &U\in(-\infty,\mu+\frac{1}{2g}]\\
        k, &U\in(\mu+\frac{k}{g}-\frac{1}{2g},\mu+\frac{k}{g}+\frac{1}{2g}]\ (k\in\Bbb N),
    \end{cases}
\end{equation}
which induces the partition $\Pi_\ast$. Working again from Definition \ref{def:PCA}, the PCA becomes
\begin{multline}
    \mathsf{PCA}(\Pi_\ast)=e^{-H}\int_{-\infty}^{\mu+\frac{1}{2g}}\frac{g}{\sigma}\phi\left(\frac{u-\mu}{\sigma/g}\right)\,\mathrm du\\
    +\sum_{k=0}^\infty\frac{e^{-H}H^k}{k!}\int_{\mu+\frac{k}{g}-\frac{1}{2g}}^{\mu+\frac{k}{g}+\frac{1}{2g}}\frac{g}{\sigma}\phi\left(\frac{u-(\mu+k/g)}{\sigma/g}\right)\,\mathrm du.
\end{multline}
After making the appropriate substitutions and simplifying, the final result is realized
\begin{equation}
    \mathsf{PCA}(\Pi_\ast)=(2-e^{-H})\Phi\left(\frac{1}{2\sigma}\right)-(1-e^{-H}).
\end{equation}
Unlike the one-bit quantizer in the previous section, because this quantizer has an infinite number of bits to encode each possible electron number, the limit of zero read noise yields $\lim_{\sigma\to 0^+}\mathsf{PCA}(\Pi_\ast)=1$.

\begin{figure}[htb]
    \centering
    \includegraphics{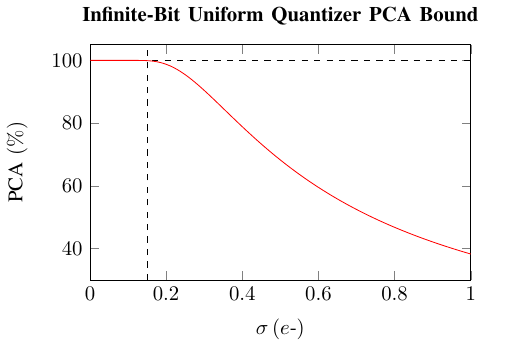}
    \caption{Lower bound of infinite-bit uniform quantizer versus read noise.  Vertical dashed line corresponds to $\sigma=0.15\,e\text{-}$.}
    \label{fig:PCA_bound}
\end{figure}

Additionally, it can be shown that $\partial_H \mathsf{PCA}(\Pi_\ast)<0$ so that taking the limit $H\to\infty$ yields a lower bound, namely,
\begin{equation}
\label{eq:PCA_lower_bound}
    \mathsf{PCA}(\Pi_\ast)\geq 2\Phi\left(\frac{1}{2\sigma}\right)-1=\operatorname{erf}\left(\frac{1}{2\sqrt 2\sigma}\right).
\end{equation}
Figure \ref{fig:PCA_bound} plots the lower bound (\ref{eq:PCA_lower_bound}) as a function of read noise.  For a read noise of $\sigma=0.15\,e\text{-}$, the lower bound is evaluated to be $\mathsf{PCA}(\Pi_\ast)|_{\sigma=0.15}=0.9991$ ($99.91\%$ accuracy).

%%%%%%%%%%%%%%%%%%%%%%%%%%%%%%%%%%%%%%%%%%%%%%%%%%%%%%%%%%%%%%%%%%%%%%%%%%%%%%%%%%%%

\subsection{Infinite-Bit Maximum Posterior Probability Quantizer}

\begin{figure}[htb]
    \centering
    \includegraphics{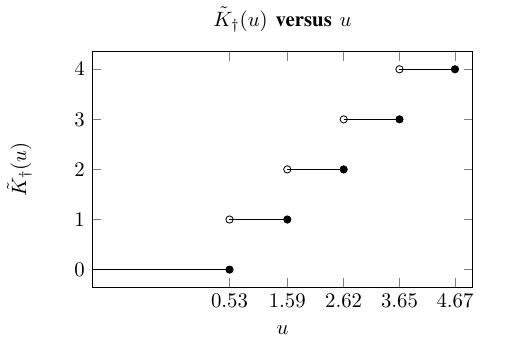}
    \caption{Infinite-bit maximum posterior probability quantizer $\tilde K_\dagger(U)$ for $H=0.75$, $g=1$, $\mu=0$, and $\sigma=0.3$. Points of discontinuity give the bin boundaries $(b_0,b_1,b_2,\dots)$.}
    \label{fig:K_versus_u}
\end{figure}

Lastly consider the infinite-bit maximum posterior probability quantizer
\begin{equation}
\begin{aligned}
    \tilde K_\dagger(U)
    &= \argmax_{k\in\Bbb N_0} \mathsf P(K=k|U=u)\\
    &= \argmax_{k\in\Bbb N_0} \frac{e^{-H}H^k}{k!}\frac{g}{\sigma}\phi\left(\frac{u-(\mu+k/g)}{\sigma/g}\right)\\
    &=
    \begin{cases}
        0, &U\in(-\infty,b_0]\\
        k, &U\in(b_{k-1},b_k]\ (k\in\Bbb N),
    \end{cases}
    \end{aligned}
\end{equation}
which induces the partition $\Pi_\dagger$. Unlike the previous examples, the partition boundaries $b_k$ cannot be written in closed-form; however, they can be computed numerically by locating the points of discontinuity of the function $\tilde K_\dagger(U)$. Substituting these numerically estimated boundaries into Definition \ref{def:PCA} subsequently allows $\mathsf{PCA}(\Pi_\dagger)$ to be numerically estimated. Figure \ref{fig:K_versus_u} plots $\tilde K_\dagger(u)$ for the parameters $H=0.75$, $g=1$, $\mu=0$, and $\sigma=0.3$ along with $x$-axis ticks marking the boundary values $(b_0,b_1,b_2,\dots)$. From the figure one can observe that the bin widths are nonuniform.

%%%%%%%%%%%%%%%%%%%%%%%%%%%%%%%%%%%%%%%%%%%%%%%%%%%%%%%%%%%%%%%%%%%%%%%%%%%%%%%%%%%%

\subsection{Numerical Example}

\begin{figure}[htb]
    \centering
    \includegraphics{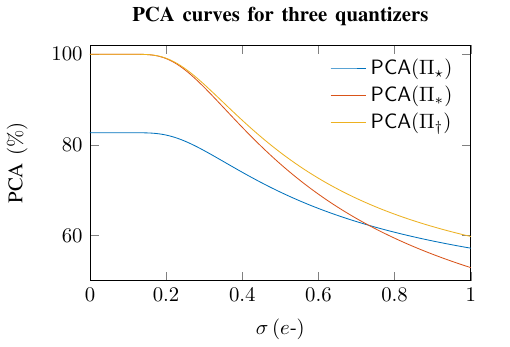}
    \caption{Photon counting accuracy versus read noise for $H=0.75$, $g=1$, and $\mu=0$.}
    \label{fig:PCA_plots}
\end{figure}

In keeping with the example in Figure \ref{fig:K_versus_u}, all three quantizers were numerically implemented for the parameters $H=0.75$, $g=1$, and $\mu=0$ and plotted as a function of read noise as seen in Figure \ref{fig:PCA_plots}. As expected, the one-bit quantizer resulted in a less than unity PCA in the limit of zero read noise, while the infinite-bit quantizers achieved unity PCA in the limit. Also of interest is the observation that the maximum posterior probability quantizer outperformed the uniform quantizer at all read noise levels. This seems reasonable as the maximum posterior probability quantizer uses information about all of the parameters $(H,g,\mu,\sigma)$ to estimate the electron number while the uniform quantizer only uses $(g,\mu)$.

%%%%%%%%%%%%%%%%%%%%%%%%%%%%%%%%%%%%%%%%%%%%%%%%%%%%%%%%%%%%%%%%%%%%%%%%%%%%%%%%%%%%
%%%%%%%%%%%%%%%%%%%%%%%%%%%%%%%%%%%%%%%%%%%%%%%%%%%%%%%%%%%%%%%%%%%%%%%%%%%%%%%%%%%%
%%%%%%%%%%%%%%%%%%%%%%%%%%%%%%%%%%%%%%%%%%%%%%%%%%%%%%%%%%%%%%%%%%%%%%%%%%%%%%%%%%%%

\section{Conclusions}

In this note a definition of photon counting accuracy was introduced as a function of a quantization strategy.  Three example quantizers were investigated and compared to determine the relative accuracy of each for a set of parameters.  This photon counting accuracy provides a simple and intuitive metric to describe the system level accuracy of modern photon counting image sensors and could find use in vendor specification sheets for these devices.

%%%%%%%%%%%%%%%%%%%%%%%%%%%%%%%%%%%%%%%%%%%%%%%%
%%%%%%%%%%%%%%%%%%%%%%%%%%%%%%%%%%%%%%%%%%%%%%%%
%%%%%%%%%%%%%%%%%%%%%%%%%%%%%%%%%%%%%%%%%%%%%%%%

\bibliographystyle{plain}
\bibliography{mybibfile}

\end{document}